\documentclass[useAMS,usenatbib]{mnras}
\usepackage{graphicx}
\usepackage{float}
\usepackage[none]{hyphenat}
\usepackage{booktabs,caption,fixltx2e}
\usepackage[flushleft]{threeparttable}
\def\simgt{\lower.5ex\hbox{$\; \buildrel > \over \sim \;$}}

\title[Molecular Outflows in HLIRGs]{Searching for molecular outflows in Hyper-Luminous Infrared Galaxies}
\author[D. Calder\'on et al.]{D. Calder\'on$^{1}$\thanks{E-mail:dcaldero@astro.puc.cl (DC)}, F.~E. Bauer$^{1,2,3}$, S. Veilleux$^{4,5}$, J. Graci\'a-Carpio$^6$, E. Sturm$^6$, 
\newauthor P. Lira$^7$, S. Schulze$^{1,2}$ and S. Kim$^1$\\
$^1$Instituto de Astrof\'isica, Facultad de F\'isica, Pontificia Universidad Cat\'olica de Chile, 782-0436 Santiago, Chile\\
$^2$Millenium Institute of Astrophysics, Santiago, Chile\\
$^3$Space Science Institute, 4750 Walnut Street, Suite 205, Boulder, Colorado 80301\\
$^4$Department of Astronomy, University of Maryland, College Park, MD 20742, USA\\
$^5$Joint Space-Science Institute, University of Maryland, College Park, MD 20742, USA\\
$^6$Max-Planck-Institut f\"ur extraterrestrische Physik, Postfach 1312, D-85741, Garching, Germany\\
$^7$Departamento de Astronom\'ia, Universidad de Chile, Casilla 36-D, Santiago, Chile}
\begin{document}

\label{firstpage}

\date{Draft \today}

\pagerange{\pageref{firstpage}--\pageref{lastpage}} \pubyear{2015}
\maketitle

\def\simlt{\lower.5ex\hbox{$\; \buildrel < \over \sim \;$}}
\def\simgt{\lower.5ex\hbox{$\; \buildrel > \over \sim \;$}}

\begin{abstract}
	We present constraints on the molecular outflows in a sample of five Hyper-Luminous Infrared Galaxies 
	using Herschel observations of the OH doublet at 119~$\mu$m. We have detected the OH doublet in three cases: 
	one purely in emission and two purely in absorption. The observed emission profile has a significant blueshifted 
	wing suggesting the possibility of tracing an outflow. Out of the two absorption profiles, one seems to be consistent 
	with the systemic velocity while the other clearly indicates the presence of a molecular outflow whose maximum velocity is 
	about $\sim$~1500~km~s$^{-1}$. Our analysis shows that this system is in general agreement with previous 
	results on Ultra-luminous Infrared Galaxies and QSOs, whose outflow velocities do not seem to correlate with 
	stellar masses or starburst luminosities (star formation rates). Instead the galaxy outflow likely arises from an 
	embedded AGN.
\end{abstract}


\section{Introduction}
	\label{sec:intro}
        Many hierarchical growth scenarios for galaxies have invoked co-evolution and symbiotic feedback between galaxies
        and their central supermassive black holes (SMBHs) \citep[e.g.,][]{M04,B06} in order to explain (1) the tight 
        relation between host galaxy spheroid and SMBH masses \citep{G00,FM00,F06,Co09}, (2) the stark contrast between 
        predicted and observed galaxy luminosity functions \citep{B03}, and (3) the crude similarity between the global 
        star formation and SMBH accretion rate histories over cosmic time \citep{M14}. In particular, it is now believed
        that strong feedback from Active Galactic Nuclei (AGNs) plays a significant role in galaxy evolution, 
        abruptly quenching star formation to produce the ``red and dead'' massive galaxies that we see today
        \citep{B03,D05,C06, B06}, and polluting the intergalactic and intracluster media with metals 
        \citep[e.g.,][]{Ga12}. Evidence for such feedback has been steadily growing over the years \citep[e.g.,][]{V05,F12,G14}. 
        For instance, detections of moderate-to-high velocity ionised outflows (100--1000 km s$^{-1}$) have now been reported 
        both as spatially resolved and kinematically distinct emission lines from ionisation regions or in absorption along 
        the line-of-sight toward numerous QSOs and Seyfert galaxies \citep[e.g.,][]{R03,K07,C07,C09,K09,R14}. These 
        studies have directly shown that AGNs can produce such feedback, and even hinted that the degree of feedback 
	may scale with SMBH mass and accretion power. Nonetheless, many questions still remain about the underlying 
        physics by which AGN-driven energy and momentum might be transferred to the interstellar medium and molecular 
        clouds ultimately responsible for star formation. 

        A number of recent studies have provided observational evidence that neutral and molecular gas outflows also appear 
        to be common in both AGNs and Ultra-Luminous Infrared Galaxies (hereafter ULIRGs). \cite{R05c} was among the first 
        to spatially resolve and map the outflow within the closest powerful ULIRG Mrk 231. \cite{R11} later detected 
        strongly blueshifted ($\sim1000$ km s$^{-1}$) optical Na\textsc{i} D $\lambda\lambda$5890, 5896 in absorption at 
        $\sim2$\,kpc from Mrk 231's nucleus, and \cite{R13b} used Keck/OSIRIS adaptive optics observations to improve the spatial resolution 
        of Mrk 231's previously observed outflow by a factor $>10$, measuring a terminal velocity of $\simgt 1300$ km\,s$^{-1}$. 
        Concurrently, \cite{Fi10} discovered a $\sim$1000 km s$^{-1}$ molecular outflow in Mrk~231 as traced by {\it Herschel} 
        far-infrared (FIR) observations of OH in absorption, while \cite{Fe10} detected spatially resolved CO emission in a 
        $\sim$~$700$~km~s$^{-1}$ outflow using the IRAM/PdB Interferometer. Remarkably, the outflow mass rate estimated from  
        OH and CO observations were in agreement within uncertainties (in the range 700--1000~M$_{\sun}$~yr$^{-1}$). 
        This estimate is significantly larger than the star formation rate (hereafter SFR) of $\sim~140$~M$_{\sun}$~yr$^{-1}$ found in the host galaxy \citep{V09}.
        Based on the energy requirements and the fact that the outflows appear to arise largely from the nucleus and not from nearby 
        star forming regions, an AGN origin is generally favored in Mrk~231.
        
        More generally, \cite{S11} reported the detection of massive molecular outflows traced by OH in several additional ULIRGs, 
	suggesting that the AGN luminosity and the AGN contribution to the total infrared luminosity seem to be correlated with 
	the outflow terminal velocities and anti-correlated with the gas depletion timescales. It is important to remark that OH lines 
	are often associated with molecular H$_2$ clouds that are somewhat perturbed from equilibrium. 
	More recently, \citet[][hereafter V13]{V13} presented a detailed study of a sample of 43 ULIRGs and QSOs at 
	$z\sim0.3$ which were observed with \textit{Herschel} using the OH doublet at 119~$\mu$m to characterise their molecular outflows. 
	V13 found that the molecular outflows in these systems do not show any obvious dependence on SFR, although they warn that 
	the range of SFRs probed by their sample is relatively narrow ($\sim1$ dex). Furthermore, V13 found that ULIRGs with 
	large AGN fractions and luminosities show OH features that are more blueshifted, particularly for objects with 
	$\log(L_{\rm AGN}/L_{\sun})\geq11.8\pm0.3$. Finally, \cite{S13} found that 15 out of 24 ULIRGs with $z<0.262$ showed evidence 
	for molecular outflows in the 79 and 119~$\mu$m OH doublets, with velocities $>700$ km s$^{-1}$, implying that AGNs were likely 
	necessary to power these high speed outflows. 
	Notably, none of these works included the most powerful luminous infrared galaxies, so-called Hyper-Luminous Infrared Galaxies 
	(hereafter HLIRGs). HLIRGs are beacons pinpointing the extremes of both star formation and perhaps AGN accretion, and thus 
	could provide clues about how the processes at work might scale with AGN power and star formation rate. Furthermore, while such 
	systems relatively rare at low redshift, ULIRGs and HLIRGs are perhaps 100 times more common at high-z \citep{C14,A15,T15}, 
	potentially providing more relevant contributions to the star formation and accretion densities at these epochs \citep[e.g.,][]{M11,M13,B15} 
	and serving as the likely progenitors for modern-day massive elliptical galaxies \citep[e.g.,][]{S88}. 
	
	In this work we analyse, for the first time, a sample of five HLIRGs in order to constrain their molecular outflows. We observed the 
	rest-frame 118--121~$\mu$m spectral region for each of these HLIRGs with  \textit{Herschel} to detect the molecular OH 
        doublet at 119~$\mu$m in emission or absorption and measure its outflow velocity. We found that three out of five HLIRGs present 
        the OH doublet feature and only one of them shows unambiguous signatures consistent with a molecular outflow. 
        In Section~\ref{sec:sample}, we describe the galaxy sample. In Section~\ref{sec:obs} we outline the reduction 
        and analysis of the observations. We present the results of this study in Section~\ref{sec:results} and 
        interpreted them in Section~\ref{sec:discussion}. Finally, we summarise our conclusions in Section~\ref{sec:conclusions}.
        In this work, we adopt $H_0=70$~km~s$^{-1}$~Mpc$^{-1}$, $\Omega_{\rm M}$ = 0.3, and $\Omega_{\Lambda}$ = 0.7. 
        We also adopt the standard convention that approaching material has a negative velocity with respect to the systemic 
        velocity of the host galaxy due to Doppler shift.
	
\section{Sample}
	\label{sec:sample}
	
	\subsection{Sample selection}
	
		Our sample was selected from a parent sample studied by \cite{R07} which contains 14 HLIRGs.
		From this sample, we took 4 relatively local galaxies  ($z<0.5$) with very high IR luminosities ($\simgt10^{13}\rm\, L_{\odot}$) 
		to build a subsample of very bright sources ($S_{100\mu \rm m}>0.7$~Jy) in order to perform high signal-to-noise ratio 
		\textit{Herschel} observations in a reasonable amount of observational time. All galaxies in the sample are late-stage mergers 
		with several lines of evidence indicating that these objects host powerful AGNs. Furthermore, we added another galaxy 
		previously observed with \textit{Herschel} and that satisfies the same selection criteria (see Section~\ref{sec:obs}).
		
		Previously measured galaxy properties are summarised in Table~\ref{tab:sample}. We initially adopted redshift 
		measurements from the literature, but found that some published values lacked associated errors and were considered less reliable. 
		To remedy this, we obtained redshifts from new spectroscopic observations for two galaxies in our sample, which we adopt below 
		and provide in Table~\ref{tab:sample}; reduction details are presented in Appendix~\ref{ap}.
		
		\begin{table*}
		\begin{threeparttable}
			\centering
			\caption{Galaxy properties}
			\begin{tabular}{cccccccc}
			\hline
			\textit{IRAS} Name	&	Type		&	$z$				&	$\log\left(\frac{L_{\rm IR}}{L_{\sun}}\right)$	&	$S_{100\mu \rm m}$  (Jy)	&	$\log\left(\frac{L_{\rm X}}{L_{\sun}}\right)$	&	t$_{\rm exp}$ (h)	&	OBS--ID\\
			\centering (1)		&	(2)		&	(3)				&	(4)							&	(5)					&	(6)							&	(7)				&	(8)\\	
			\hline
			\hline
			F00183--7111$^*$	&	AGN2	&	$0.3282\pm0.0005$	&	13.13						&	1.19					&	11.6							&	6.7				&	1342245966\\  	
			00397--1312		&	AGN2	&	$0.2617\pm0.0001$	&	12.90						&	1.90					&	--							&	4.9				&	1342238350\\
			07380--2342		&	AGN2	&	$0.2924\pm0.0002$	&	13.49						&	3.55					&	$<9.0$						&	0.8				&	1342245974\\	
			12514+1027$^*$	&	AGN2	&	$0.3192\pm0.0003$	&	12.92						&	0.76					&	$10.0$						&	11.8				&	1342248537, 1342248539\\ 	
			14026+4341		&	QSO1	&	$0.3243\pm0.0003$	&	13.11						&	0.99					&	$<9.3$						&	9.6				&	1342257687, 1342257688\\  	
			\hline
			\end{tabular}
			\label{tab:sample}
			\begin{tablenotes}
				\item	\textbf{Notes.} Column 1: \textit{IRAS} galaxy names. Column 2: AGN type from \cite{R07}. 
				Column 3: redshifts from \cite{S00}, except for the objects marked with an asterisk for which we used 
				our own estimated values (see Appendix~\ref{ap}). Column 4: IR luminosity $L_{\rm IR}$, spanning 
				8--1000~$\mu$m \cite{R07}. Column 5: flux density at 100 $\mu$m from \cite{R07}. Column 6: X-ray 
				luminosity $L_{\rm X}$ spanning 0.2--10~keV from \cite{R07}. Column 7: total PACS exposure time of 
				our observations, except for IRAS 00397--1312 which was previously observed as part of programme 
				\textit{OT1\_dfarrah\_1}. Column 8: \textit{Herschel} observation ID. 
			 \end{tablenotes}
		\end{threeparttable}
		\end{table*}

	\subsection{AGN fraction and stellar masses}
		
		We proceed to characterise the AGN power of the galaxies in our sample, based on 
		the estimated galaxy properties from the previous subsection. Firstly, we calculated 
		the bolometric luminosity $L_{\rm bol}$ of the galaxies between 8--1000~$\mu$m, 
		the so-called IR luminosity $L_{\rm IR}$ \citep{S96}, which are listed in Table~\ref{tab:sample}. 
		To accomplish this, we used the expression $L_{\rm bol}=1.15L_{\rm IR}$ from \cite{V09}, 
		based on the average value for local ULIRGs \citep{K98}. Then, we calculated the so-called 
		\textit{AGN fraction}, which is the fractional contribution of the AGN to the total bolometric luminosity, i.e.
		 
		 \begin{eqnarray}
			L_{\rm bol}	&	=	&	L_{\rm AGN}	+	L_{\rm SB},\label{eq:agn}\\
						&	=	&	\alpha_{\rm AGN}L_{\rm bol}	+	L_{\rm SB}\label{eq:sb},
		\end{eqnarray}
		
		\noindent where $L_{\rm AGN}$ and $L_{\rm SB}$ are the AGN and starburst contributions to the total luminosity of a given galaxy, 
		respectively. \cite{V09} showed that the rest-frame 15 to 30~$\mu$m continuum ratio was correlated with the PAH-free, 
		silicate-free mid-IR (MIR)/FIR ratio and the AGN contribution to the bolometric luminosity more than any other \textit{Spitzer}-derived 
		continuum ratio they explored; we adopt the prescription based on this finding, which is referred as \textit{Method \#6} in their work. 
		Basically, it consists in adopting $\log(f_{30}/f_{15})=0$ and 1.35 as zero points for pure AGN and pure starburst ULIRGs, respectively 
		(entries in Table 9 from \citealt{V09}). Then, using the bolometric corrections from Table 10 and the formula given in 
		the notes of Table 11, both from \cite{V09}; we obtain the bolometric $\alpha_{\rm AGN}$. 
		To apply this procedure, we make used of \textit{Spitzer} spectra in the range $\sim4$--40 $\mu$m of the galaxies in our sample, which 
		we obtained from the \textit{IRS Enhanced Products} catalogue. As none of the rest-frame spectra sampled the continuum at 30 um 
		(except for IRAS 00397--1312), we linearly extrapolated the data we had to estimate a value of the continuum at this wavelength. 
		The absolute uncertainty on $\alpha_{\rm AGN}$ values is about $\pm20$\% on average. With this estimate and the bolometric 
		luminosity we can compute $L_{\rm AGN}$ and $L_{SB}$ using Equations~\ref{eq:agn} and~\ref{eq:sb}. 
		Furthermore, we measured the depth of the silicate feature at 9.7~$\mu$m, $\rm Si_{\rm 9.7\mu m}$ \citep{St13}. This quantity is a measure of 
		the obscuration in ULIRGs. This depth is estimated as the logarithm of the ratio of the measured flux at 9.7~$\mu$m relative to the local 
		continuum (i.e., more negative values indicate stronger absorption features). The results of this analysis are shown in Table~\ref{tab:lum}.
		
		\begin{table*}
		\begin{threeparttable}
			\centering
			\caption{Estimated galaxy properties}
			\begin{tabular}{ccccccc}
				\hline
				\textit{IRAS} Name	&	$f_{15}/f_{30}$	&	$\alpha_{\rm AGN}$ (\%)	&	$\log\left(L_{\rm bol}/L_{\sun}\right)$	&	$\log\left(L_{\rm bol}^{\rm AGN}/L_{\sun}\right)$	&	$\log\left(L_{\rm bol}^{\rm SB}/L_{\sun}\right)$\	&	Si$_{9.7 \mu \rm m}$\\
				(1)			&	(2)		&	(3)		&	(4)		&	(5)		&	(6)		&	(7)\\	
				\hline
				\hline
				F00183--7111	&	0.117	&	52		&	13.19	&	12.91	&	12.87	&	$-1.13\pm0.02$	\\
				00397--1312	&	0.152	&	63		&	12.96	&	12.76	&	12.53	&	$-1.09\pm0.03$	\\
				07380--2342	&	0.855	&	99		&	13.55	&	13.54	&	11.66	&	$-0.08\pm0.01$	\\
				12514+1027	&	0.270	&	80		&	12.98	&	12.89	&	12.27	&	$-0.61\pm0.02$	\\
				14026+4341	&	0.420	&	90		&	13.17	&	13.12	&	12.19	&	$-0.09\pm0.01$	\\
				\hline
			\end{tabular}
			\label{tab:lum}
			\begin{tablenotes}
				\item	\textbf{Notes.} {\it Column 1:} \textit{IRAS} galaxy name. {\it Column 2:} 15-to-30 $\mu$m flux ratio estimated from the \textit{Spitzer} spectra. 
				{\it Column3}: Percentage AGN contribution to the bolometric luminosity based on the $f_{15}/f_{30}$ method. {\it Column 4:} Bolometric luminosity obtained 
				from the IR luminosity following \cite{V09}. {\it Column 5:}  AGN bolometric luminosity obtained from the total bolometric luminosity and the AGN fraction. {\it Column 6:} 
				Starburst bolometric luminosity obtained from the total bolometric luminosity and the AGN fraction. {\it Column 7:} Depth of the silicate absorption 
				feature at 9.7~$\mu$m estimated following \cite{St13} using the \textit{Spitzer} spectral data.
			 \end{tablenotes}
		\end{threeparttable}
		\end{table*}
		
		Finally, we made use of photometric measurements as tracers of the stellar mass content in the galaxies.
		We used \textit{HST} images obtained from the Mikulski Archive database\footnote{https://archive.stsci.edu/} 
		and extracted the central point source from each galaxy, in order to exclude the AGN contribution, before performing the photometry.
		To do so, we generated theoretical PSFs using \textit{Tiny Tim} \citep{K11} and then we subtracted them from the extended sources using \textit{GALFIT} \citep{P02}. 
		Finally, we ran \textit{Source Extractor} \citep{B96} on the residual images and estimated Vega magnitudes to make direct comparisons with 
		results from V13. Unfortunately, we found \textit{HST} imaging data for only four of the five galaxies in our sample: two with H-band (\textit{F160W}) 
		and two with I-band (\textit{F814W}). To augment these, we include H-band data for three sources from the 2MASS data archive, which were not PSF 
		subtracted. The 2MASS magnitudes at least provide strict upper limits on the stellar mass. The errors in the \textit{HST} photometry are $\sim0.5$ mag, 
		incorporating both the PSF subtraction and photometry errors. The results of this procedure are presented in Table~\ref{tab:phot}.
		
		\begin{table*}
			\begin{threeparttable}
			\centering
			\caption{Galaxy photometric measurements}
			\begin{tabular}{ccccc}
				\hline
				IRAS Name	&	$m_{\rm F160W}$			&	$m_{\rm F814W}$	&	$M_{\rm F160W}$			&	$M_{\rm F814W}$	\\
				(1)			&	(2)						&	(3)				&	(4)						&	(5)\\
				\hline
				\hline
				F00183--7111	&	$15.92\pm0.18^{\rm a}$	&	-				&	$-24.81\pm0.18^{\rm a}$	&	-			\\
				00397--1312	&	$16.6\pm0.5$				&	-				&	$-23.6\pm0.5$				&	-			\\
				07380--2342	&	$14.70\pm0.07^{\rm a}$	&	$17.3\pm0.5$		&	$-25.80\pm0.07^{\rm a}$	&	$-23.2\pm0.5$	\\					
				12514+1027	&	$15.21\pm0.11^{\rm a}$	&	$18.6\pm0.5$		&	$-22.48\pm0.11^{\rm a}$	&	$-22.1\pm0.5$	\\
				14026+4341	&	$>18.8$					&	-				&	$>-21.9$					&	-			\\
				\hline
			\end{tabular}
			\label{tab:phot}
			\begin{tablenotes}
				\item	\textbf{Notes.} {\it Column 1:} \textit{IRAS} galaxy names. {\it Column 2:} PSF-subtracted apparent magnitude in the \textit{HST F160W} band. 
				{\it Column 3:} PSF-subtracted apparent magnitude in the \textit{HST F814W} band. {\it Column 4:} Absolute magnitude in the \textit{HST F160W} band. 
				{\it Column 5:} Absolute magnitude in the \textit{HST F814W} band.
				\item[$^{\rm a}$] H-band magnitudes without performing PSF subtraction taken from the 2MASS data archive.
			 \end{tablenotes}
		\end{threeparttable}
		\end{table*}
		
	\section{Observations, Data Reduction and Spectral Analysis}
		\label{sec:obs}
		
		\subsection{Observations}
		
			The HLIRG observations  were carried out with the Photodetecting Array Camera and Spectrometer 
			\citep[PACS;][]{Po10} installed on \textit{Herschel} \citep{Pi10}. 
			 Four galaxies were observed as part of programme \textit{OT2\_fbauer\_1}, while 
			 the fifth object (IRAS~00397--1312) of our sample was previously observed also with \textit{Herschel}/PACS 
			 as part of programme \textit{OT1\_dfarrah\_1}.
			We used PACS in the range scan spectroscopy mode with high-sampling, centered on the redshifted 
			OH~119~$\mu$m~$+$$^{18}$OH~120~$\mu$m complex with a velocity range of $\sim4000$ km s$^{-1}$ 
			(rest-frame 118--121 $\mu$m), which is needed to provide enough spectral coverage on both sides of the 
			OH complex for reliable continuum assessment. The resolution for our targets was $\sim240$km~s$^{-1}$, 
			which allowed us to detect outflow velocities up to 1000--2000~kms$^{-1}$.
		
		\subsection{Data reduction}
		
			The reduction procedure was identical to the one performed in \cite{S11} and V13. A standard PACS reduction and calibration was 
			applied using the pipeline {\em ipipe} from HIPE 6.0. The spectra were normalised to the telescope flux and then recalibrated with previous 
			Neptune observations during \textit{Herschel}'s performance verification phase. 
			Although the sources were not point-source corrected, this fact would not change the spectral shape of the sources. 
			In all that follows, we only use the central spatial pixel alone due to its better S/N. 
			For two objects in the sample (12514+1027 and 14026+4341) we obtained two separate observations. Therefore we co-added them 
			after confirming that they were consistent with each other. To combine them, we averaged the two observations as 
			both exposure times were basically the same. After doing so, we analysed only the combined spectra.
			
		\subsection{Spectral analysis}
		
			Our analysis is analogous to the one performed by V13 on their ULIRGs+QSOs sample, which we describe briefly. The reduced \textit{Herschel} 
			spectra were smoothed using a Gaussian kernel with $\sigma = 0.025$ $\mu$m to reduce the noise before analysing them. Then we fitted 
			the base continuum level with a simple zero--order polynomial and divided the spectra by their continuum, in order to analyse the significance of the OH detection. 
			When there was obvious curvature or a non-zero slope in the continuum component, we fitted a spline function instead. 
			To characterise the OH~119.233, 119.441 $\mu$m doublet, we used two velocity components (a total of four Gaussians for the OH doublet). 
			When the fit statistic for a single velocity component model was equal to or better than that of a two-velocity component model, we favored the former.
			The least-square fits used the Levenberg-Marquardt algorithm to model the observed lines. Each Gaussian component is defined by its amplitude, peak 
			position and standard deviation (FWHM). We kept fixed the separation of the lines to 0.208 $\mu$m in the rest-frame ($\sim520$~km~s$^{-1}$) and required 
			that the amplitude and the standard deviation be fixed to the same value for each component in the doublet.
			
			We analysed the OH doublet profile by measuring characteristic velocities from our fits, i.e., both velocity components fitted to the OH doublet; 
			(1) $v_{50}$(abs), the median velocity of the fitted absorption profile, (2) $v_{84}$(abs), the velocity above which 84\% of the absorption takes place, 
			(3) $v_{50}$(emi), the median velocity of the fitted emission profile, and $v_{84}$(emi), the velocity below which 84\% of the emission takes place. 
			Also, we included the terminal outflow velocity, $v_{\rm max}$, in our analysis, which is obtained from the maximum extent of the blueshifted wing 
			of the 119.233 $\mu$m profile. The uncertainties on these characteristic velocities measurements are dominated by where we placed the continuum, 
			and therefore we estimated errors by assuming different continuum shapes (spline or polynomial) and their positions. We found errors in $v_{50}$ of 
			$<100$ km s$^{-1}$, in $v_{84}$ of $\sim100$ km s$^{-1}$ and in $v_{\rm max}$ of $\sim200$ km s$^{-1}$.

	\section{Results}
	\label{sec:results}
		The results of the OH 119~$\mu$m profile fits are shown in Figures~\ref{fig:00183}--\ref{fig:14026}, while the parameters 
		derived from the spectral analysis are listed in Table~\ref{tab:vel}. The OH doublet was detected in three 
		out of five galaxies: in two galaxies (IRAS F00183-7111 and 12514+1027), the detections were purely in absorption, while in 
		another (IRAS 00397--1312) it is purely in emission. For the two other objects (IRAS 07380--2342 and 14026+4341), 
		we find that the spectra appear to be composed solely of continuum emission, and therefore the OH doublet may be either too weak or simply absent.
		
		\subsection{Source by source analysis}
		
		\subsubsection{IRAS F00183-7111}
		
\begin{figure*}
			\centering
			\includegraphics[width=0.495\textwidth]{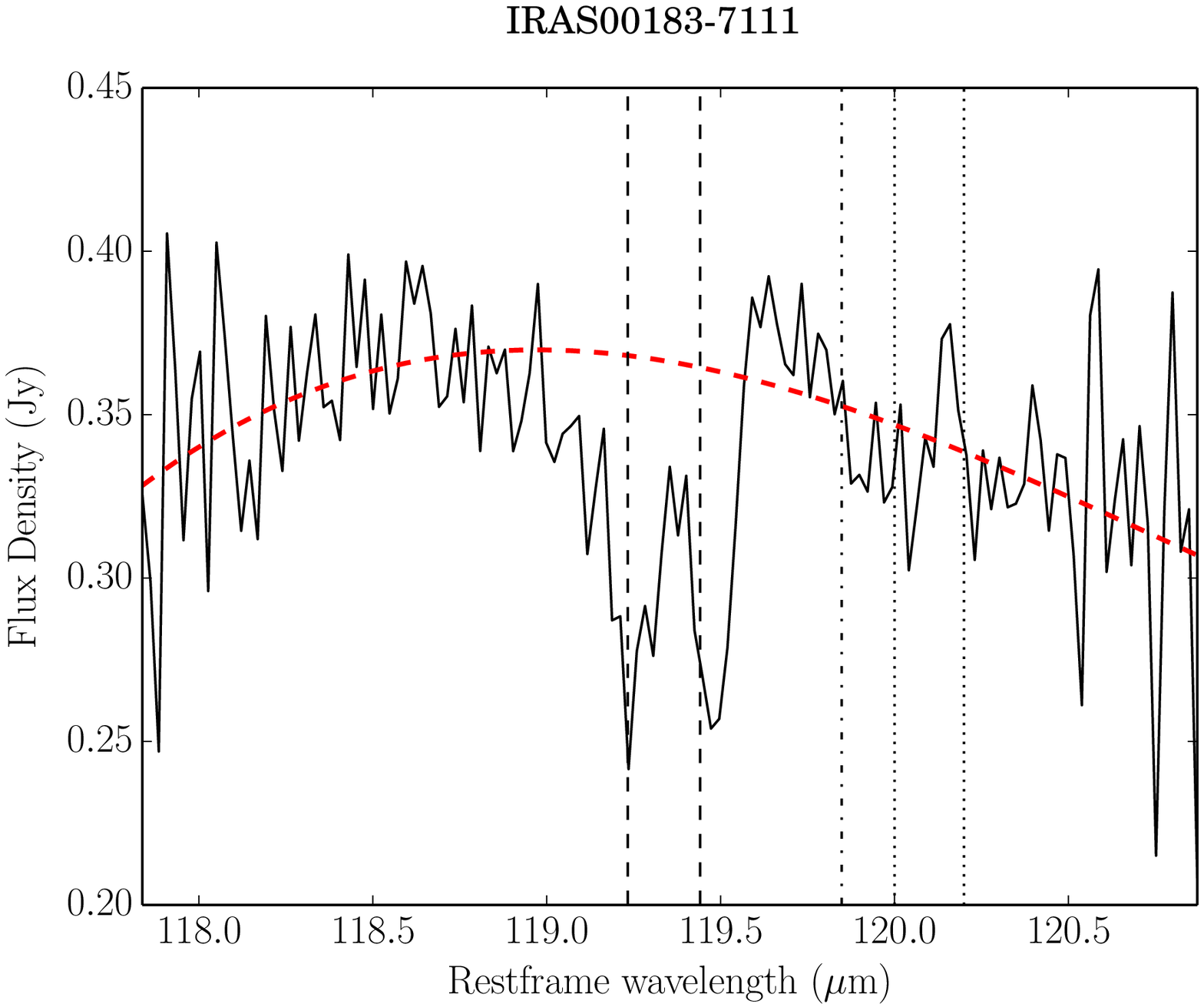}
			\includegraphics[width=0.495\textwidth]{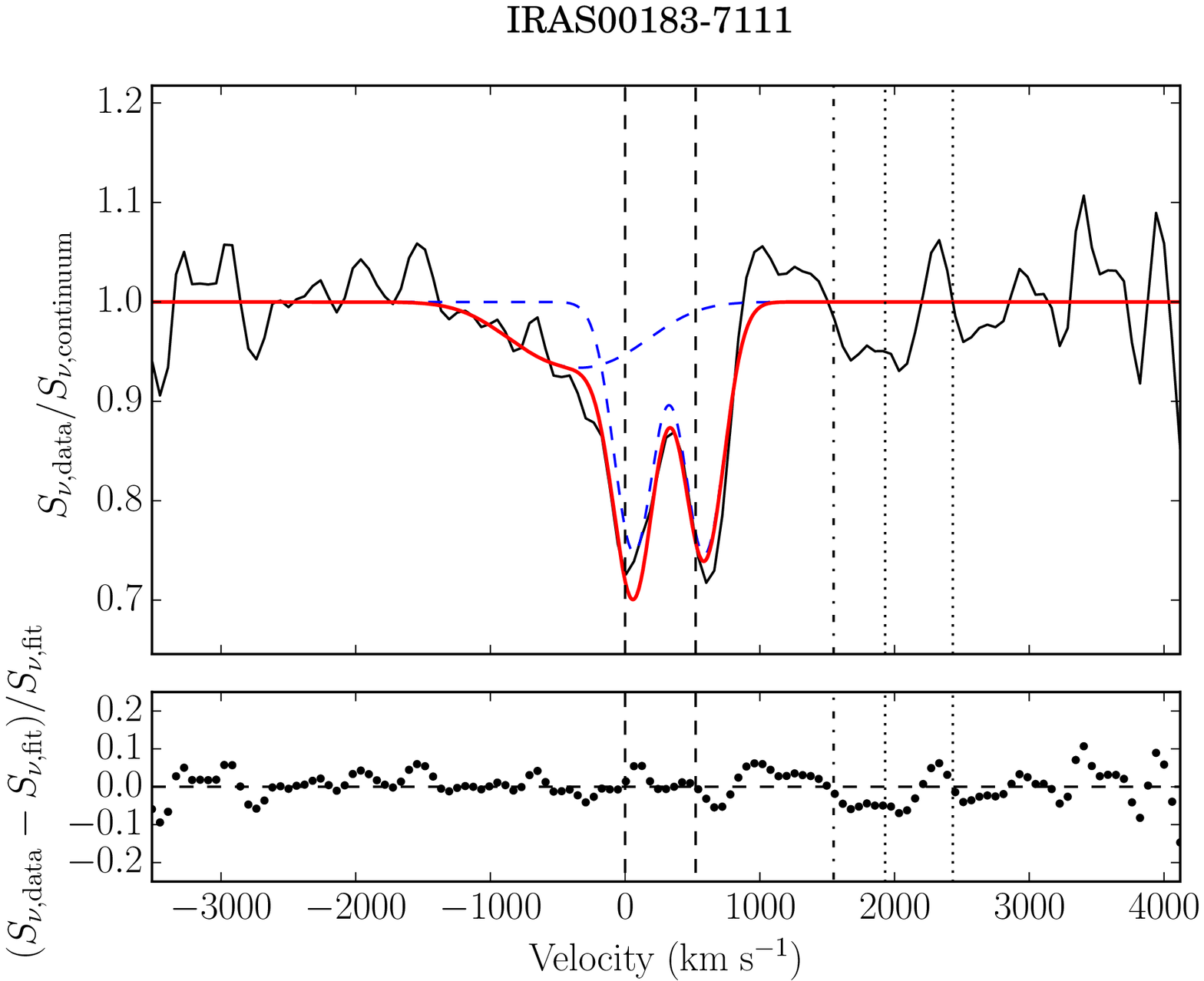}
			\caption{Spectral fits to the OH 119 $\mu$m doublet for the objects in our sample. 
			On the left panel it is shown the reduced, not point-source corrected spectrum of IRAS F00183-7111 
			where the dashed red line represents the continuum fitted. On the right panel it is shown the smoothed spectrum 
			divided by the continuum fit (solid black line) where the dashed blue lines are the velocity components 
			of the doublet fit and the solid red line represents the total doublet fit (i.e., the sum of the components).
			The origin of the velocity in the x-axis corresponds to OH $119.233\,\mu$m at the systemic velocity. 
			The two vertical dashed and dotted lines stand for the restframe location of the OH $119\,\mu$m 
			and $120\,\mu$m doublets, respectively. The vertical dot-dash line shows the position of the CH$^+$ $119.848\,\mu$m line.
			Furthermore, the lower box of the right panel shows the residuals of the fit on the smoothed data.}
			\label{fig:00183}
		\end{figure*}

			On the left side panel of Figure~\ref{fig:00183}, we show the spectrum before analysing it. 
			We can clearly see the OH doublet feature and probably a weak emission component. 
			From the analysis, the best fit was reached by fitting a spline function to the continuum, 
			which is denoted by a dashed red line. The right panel shows the spectrum after being smoothed 
			and divided by the fitted continuum. This allows us to distinguish that there is a significant blueshifted wing 
			beyond the instrumental Gaussian tail of the strongest absorption component which we interpret as 
			a molecular outflow with a $v_{\rm max}\sim 1500\rm\, km\ s^{-1}$.
			
		\subsubsection{IRAS 00397--1312}
		
\begin{figure*}
			\centering
			\includegraphics[width=0.495\textwidth]{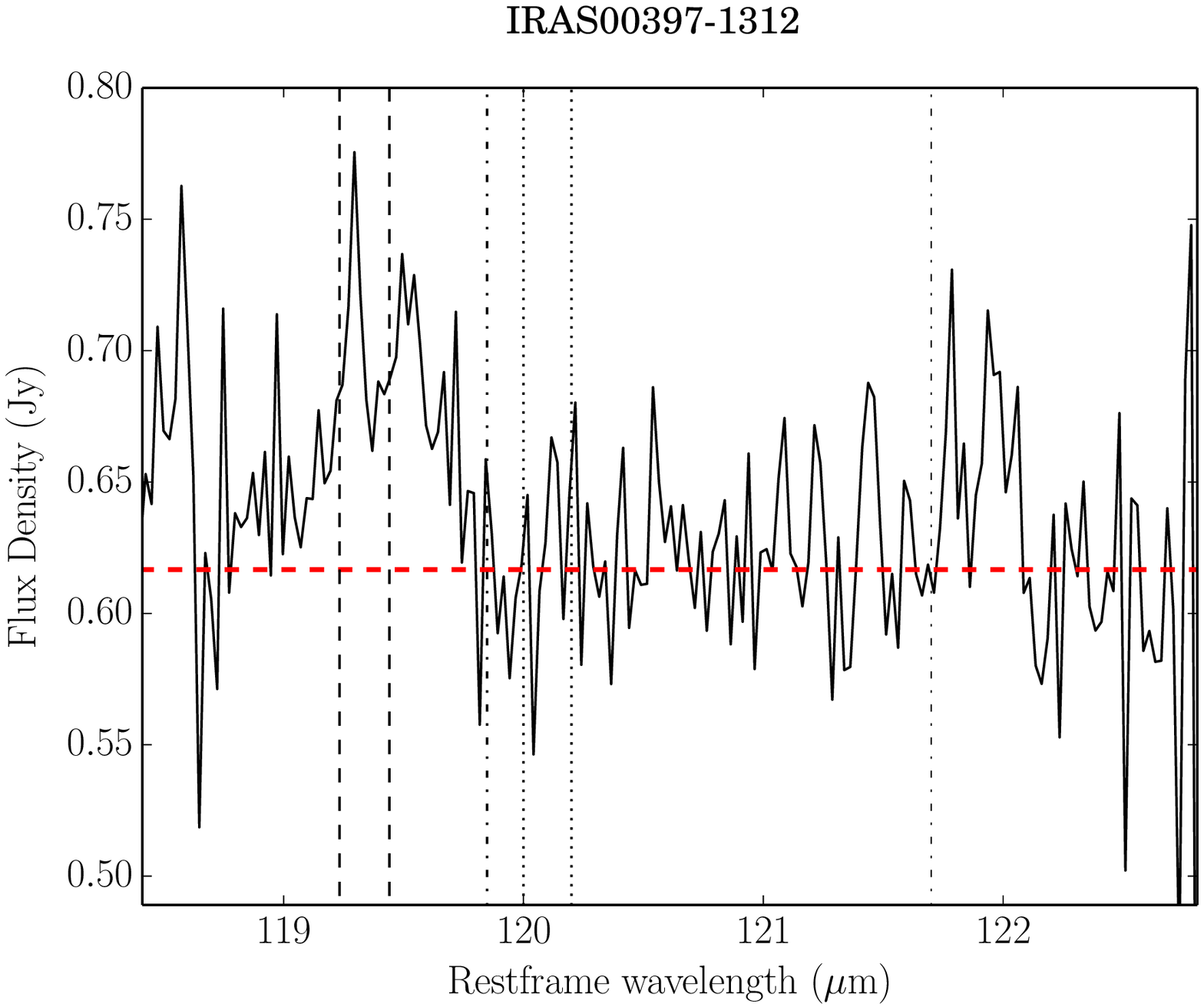}
			\includegraphics[width=0.495\textwidth]{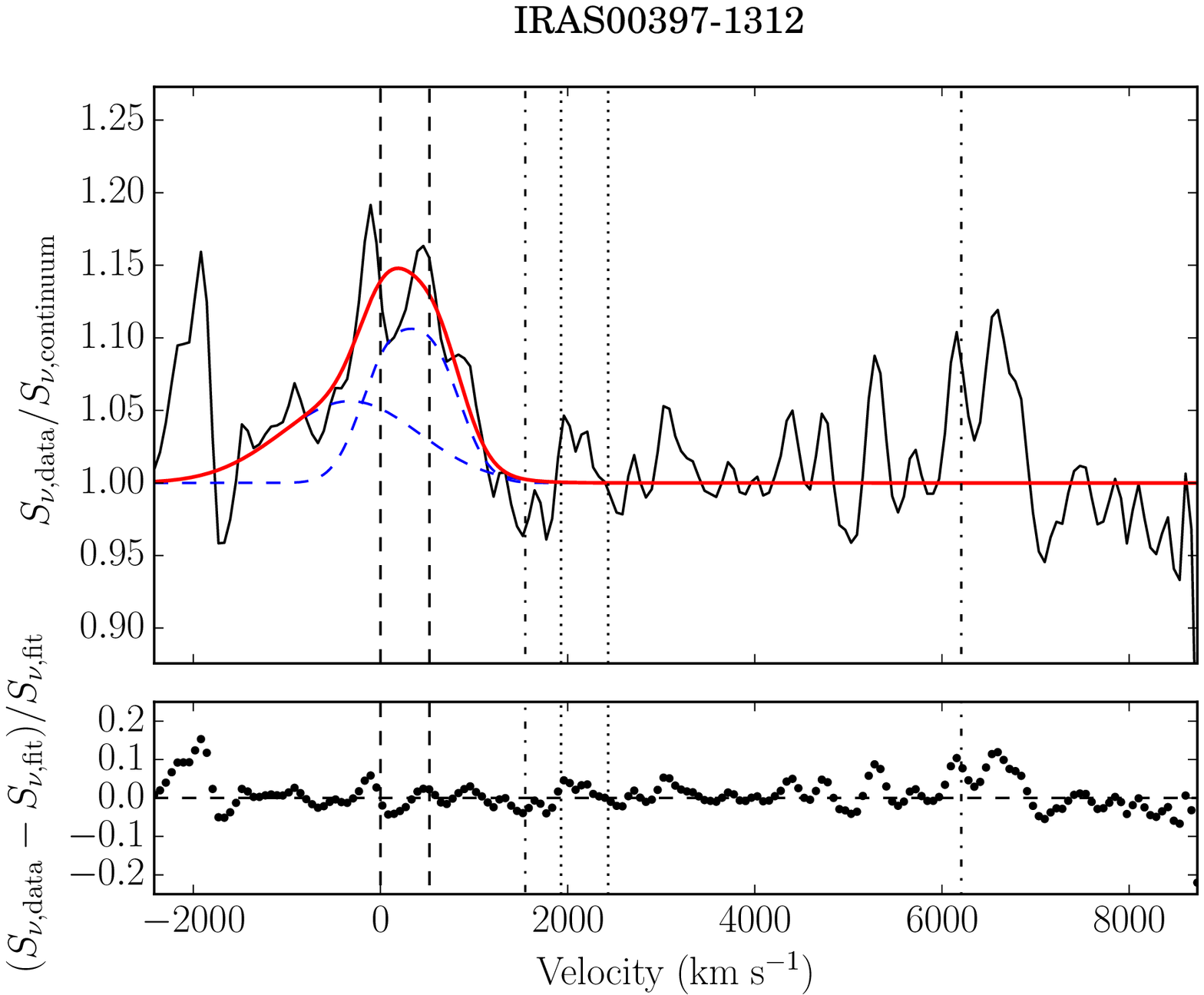}
			\caption{Analogous to Figure~\ref{fig:00183} but for the galaxy IRAS 00397--1312. 
			The vertical dot-dash line here at $\lambda\sim121.7\,\mu$m ($v\sim6100\,$km~s$^{-1}$) shows the position of the [N\textsc{ii}] line.}
			\label{fig:00397}
		\end{figure*}

			The spectrum is shown in Figure~\ref{fig:00397} where we can recognise the OH doublet 
			in emission and a weaker [NII]121.7 fine-structure line around $\lambda\sim122\rm\, \mu m$. 
			In this case, a better fit was found using a flat slope for the continuum. On the right 
			panel, we show the spectrum fitting the OH doublet with two doublets. A second doublet 
			is needed to take into account the blueshifted wing that departs from the primary component. 
			This feature may indicate the presence of an outflow, however without an absorption component we cannot be certain.
			Furthermore, a possible redshifted [NII] emission line may be present, although the signal-to-noise 
			is too low to characterise it well.
		
		\subsubsection{IRAS 07380--2342}
		
\begin{figure*}
			\centering
			\includegraphics[width=0.495\textwidth]{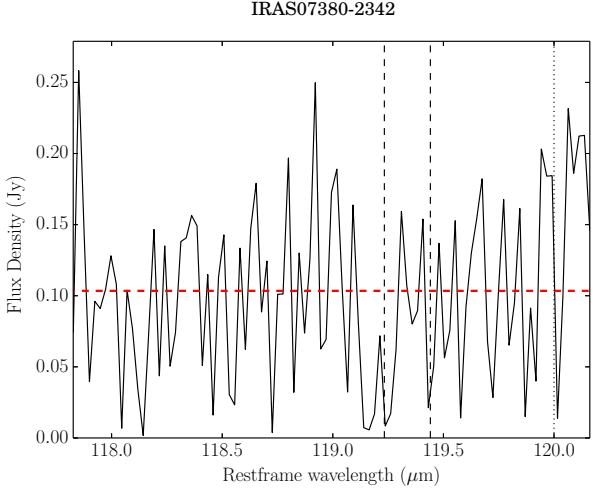}
			\includegraphics[width=0.495\textwidth]{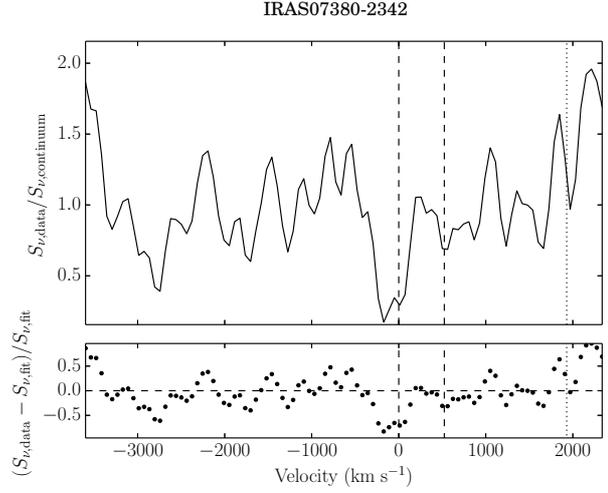}
			\caption{Analogous to Figure~\ref{fig:00183} but for the galaxy IRAS 07380--2342.}
			\label{fig:07380}
		\end{figure*}

			The spectrum of this galaxy is shown in the left panel of Figure~\ref{fig:07380}. 
			Although the doublet might be detected marginally in absorption, it is difficult to infer 
			its presence with certainty due to the fact that the spectrum seems to be heavily dominated by the noise. 
			This might be caused by a pointing offset when observing this galaxy, as the brightest pixel of the image is not at the centre.
		
		\subsubsection{IRAS 12514+1027}
		
\begin{figure*}
			\centering
			\includegraphics[width=0.495\textwidth]{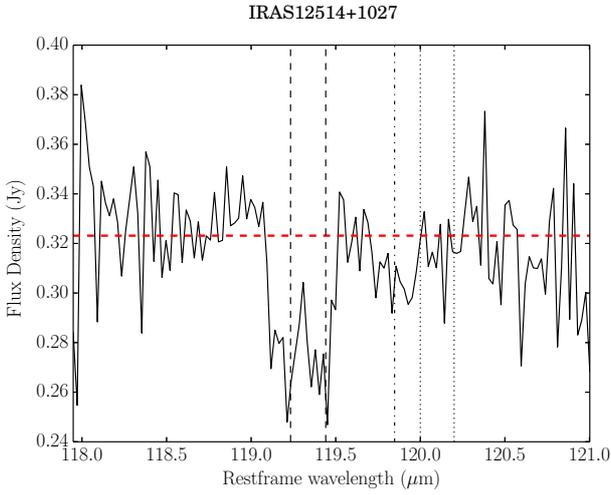}
			\includegraphics[width=0.495\textwidth]{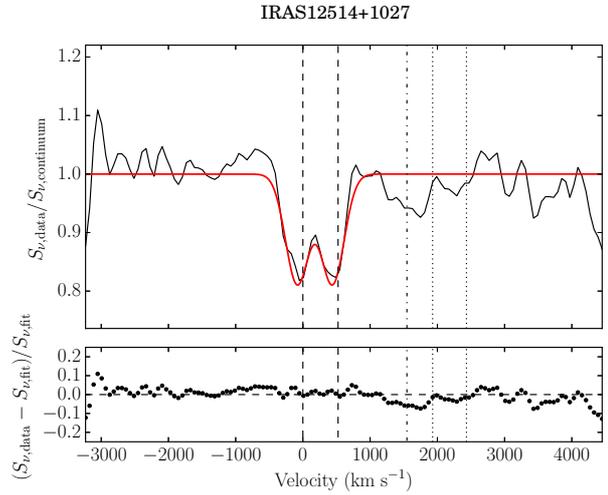}
			\caption{Analogous to Figure~\ref{fig:00183} but for the galaxy IRAS 12514+1027. }
			\label{fig:12514}
		\end{figure*}

			On the left panel of Figure~\ref{fig:12514}, we see prominent absorption lines which 
			correspond to the OH doublet and a weaker feature which could indicate the presence of the 
			$^{18}$OH 120$\,\mu$m doublet. Although we only used a single velocity component to fit the 
			OH 119$\,\mu$m doublet, it is possible that a bump at $-1000$~km~s$^{-1}$ is also part of the doublet, 
			implying more complex velocity structure. However, as the main doublet is consistent with the systemic 
			velocity of the galaxy, we cannot confirm a molecular outflow detection from this spectrum (see right panel of Figure~\ref{fig:12514}).

		\subsubsection{IRAS 14026+4341}
		
\begin{figure*}
			\centering
			\includegraphics[width=0.495\textwidth]{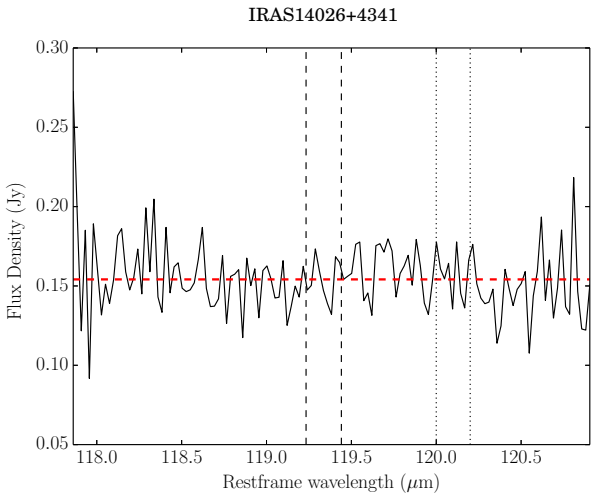}
			\includegraphics[width=0.495\textwidth]{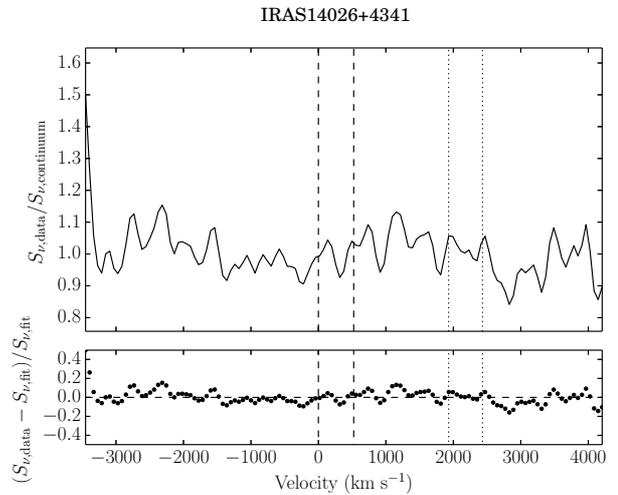}
			\caption{Analogous to Figure~\ref{fig:00183} but for the galaxy IRAS 14026+4341.}
			\label{fig:14026}
		\end{figure*}

			In this case the spectrum appears only to be composed of continuum emission and is consistent 
			with a flat model component(see Figure~\ref{fig:14026}).
		
	\section{Discussion}
	\label{sec:discussion}	
\begin{figure*}
		   \centering
		   \includegraphics[width=1.05\linewidth]{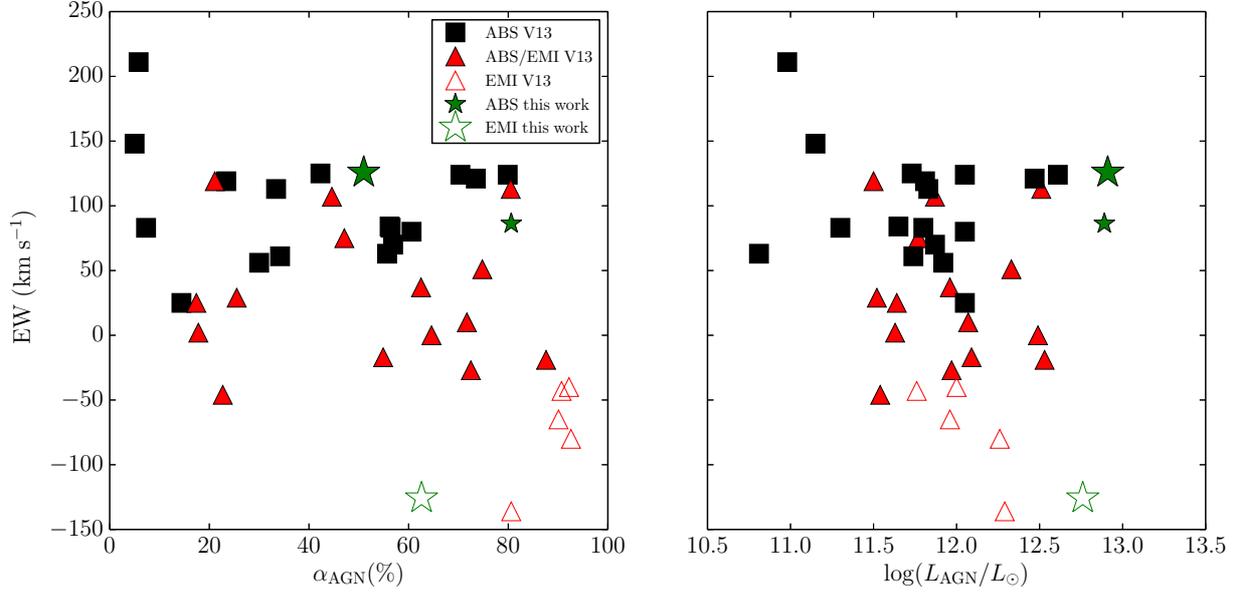} 
		   \caption{Total equivalent width of the 119 $\mu$m doublet as a function of 
		   AGN fraction (left panel) and AGN luminosity (right panel). Black squares, 
		   filled red triangles, and open red triangles are ULIRGs from V13 with OH 
		   119 $\mu$m seen purely in absorption, composite absorption/emission, and purely 
		   in emission, respectively. Meanwhile, filled green and open green stars
		   are HLIRGs from this work with the OH doublet seen purely in absorption and purely in emission.}
		   \label{fig:ew}
		\end{figure*}

\begin{figure*}
		   \centering
		   \includegraphics[width=0.6\linewidth]{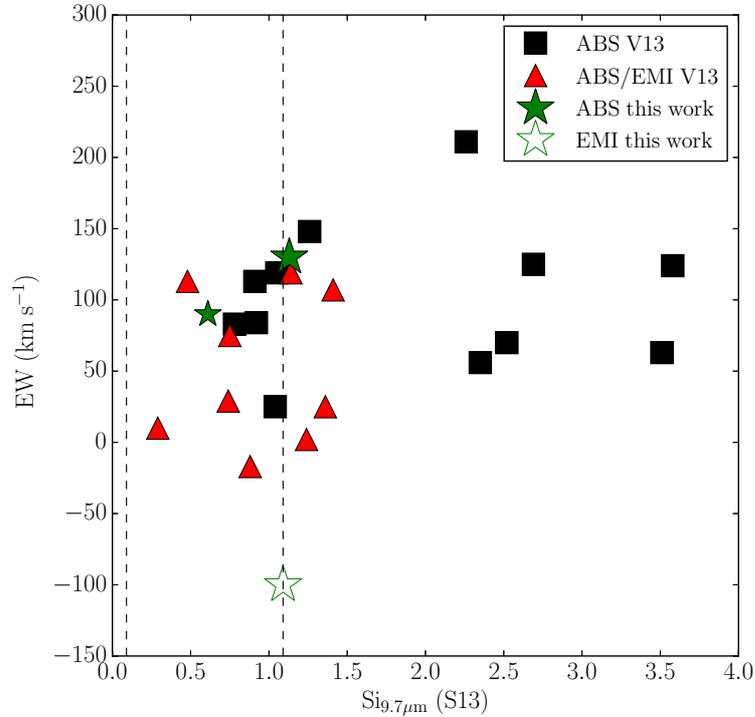} 
		   \caption{Equivalent width of the 119 $\mu$m doublet as a function of 
		   the depth of the silicate feature at 9.7~$\mu$m relative to the local continuum 
		   on a logarithmic scale measured by \protect\cite{St13}. The meaning of the symbols 
		   is the same as in Figure~\ref{fig:ew}. Vertical black dashed lines stand for the HLIRGs 
		   without OH detection.}
		   \label{fig:si}
		\end{figure*}

\begin{figure*}
		   \centering
		   \includegraphics[width=0.9\linewidth]{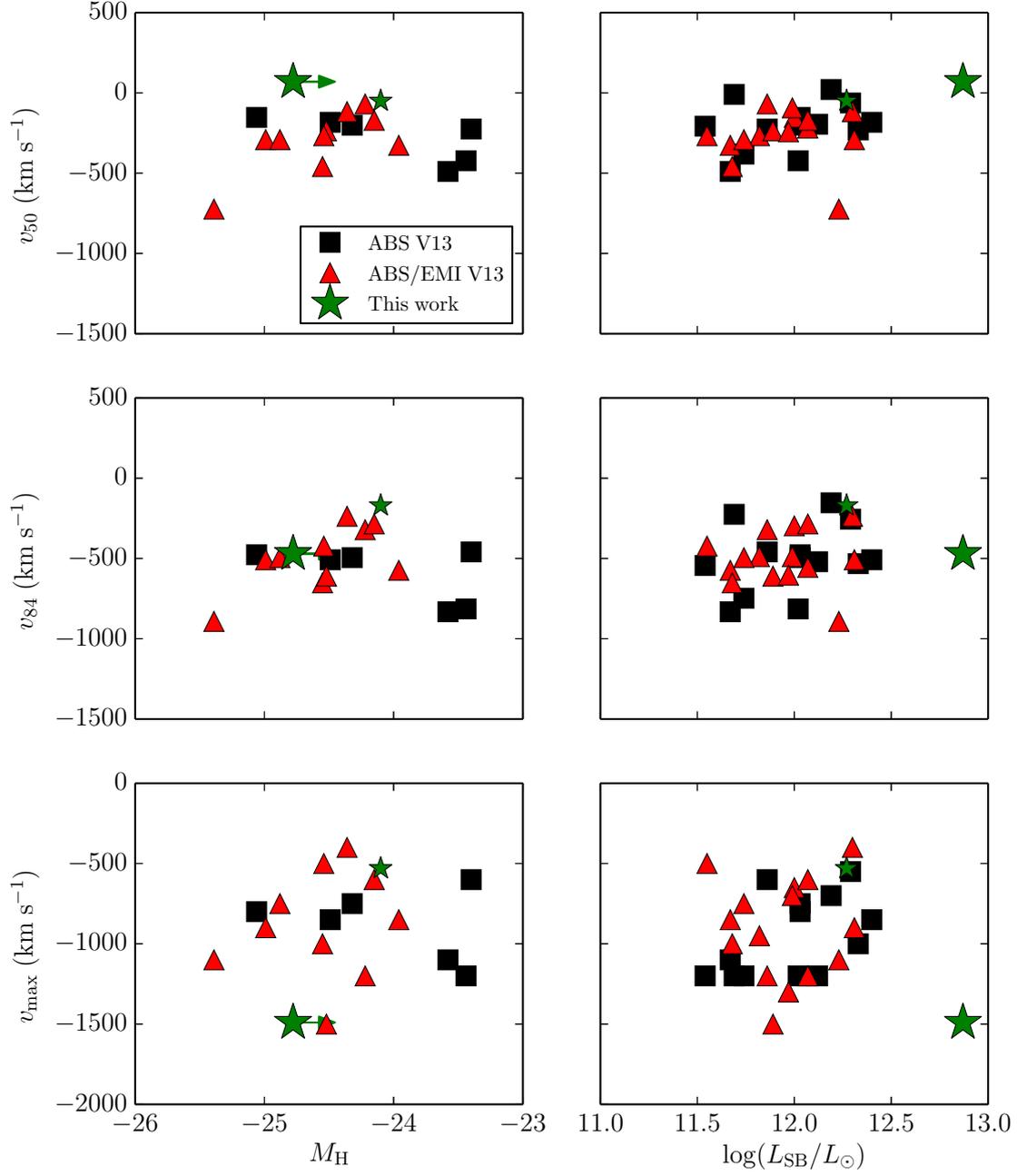} 
		   \caption{50\% (upper), 84\% (central), and terminal (lower) 
		   OH outflow as a function of the absolute magnitude in H-band on the left panels, and 
		   starburst luminosity on the right panels. The meaning of the symbols is the same as in Figure~\ref{fig:ew}.}
		   \label{fig:sb}
		\end{figure*}

\begin{figure*}
		   \centering
		   \includegraphics[width=0.9\linewidth]{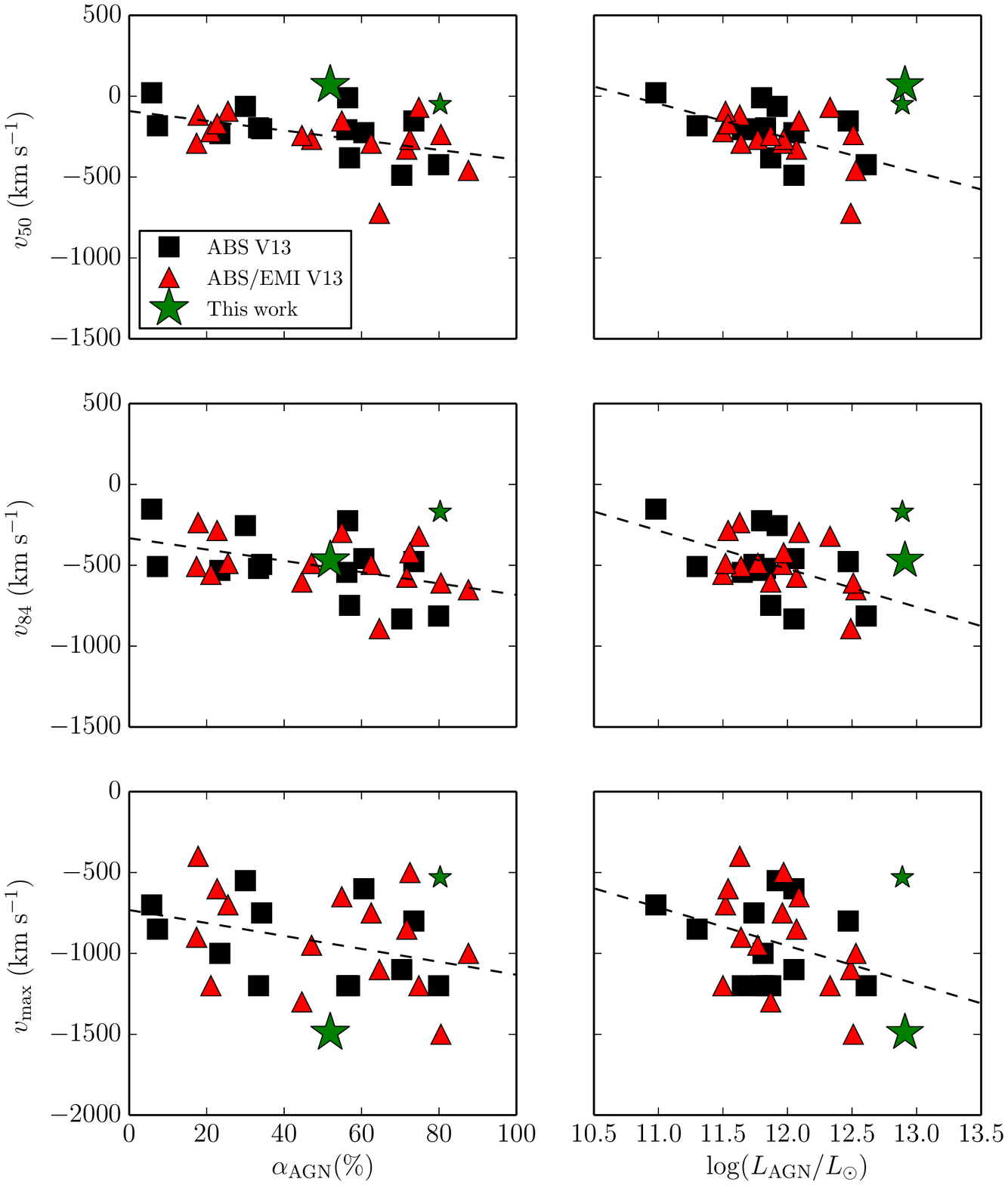} 
		   \caption{50\% (upper), 84\% (central), and terminal (lower) 
		   OH outflow as a function of the AGN fractions on the left panels, and 
		   AGN luminosity on the right panels. The meaning of the symbols is the same as in Figure~\ref{fig:ew}.}
		   \label{fig:agn}
		\end{figure*}

		We compared the spectral measurements with derived galaxy properties from Tables~\ref{tab:lum} and~\ref{tab:phot} 
		to look for any signs of correlations. We included here the data from V13 (shown as black squares and red triangles), 
		who performed an analogous procedure in a larger ULIRG sample. In this analysis we included the three HLIRGs with 
		OH detections, however we have to keep in mind that only one source has signatures of having a molecular outflow. 
		To highlight this fact, this galaxy is shown as a big green filled star in Figures~\ref{fig:ew}--\ref{fig:agn}.
		
		Figure~\ref{fig:ew} shows a general lack of correlation between the equivalent width of the OH feature 
		and AGN fraction or AGN luminosity found by V13. Furthermore, we see our data points do not change this result.
		We have also studied the relation between the OH equivalent width and the obscuration of the systems using Si$_{\rm 9.7\mu m}$.
		This comparison is shown in Figure~\ref{fig:si}, which plots the sources from V13 and this work data. 
		The HLIRGs in our sample are not very obscured and do not have strong OH absorption, which principle supports the correlation 
		found by V13 that more obscured objects present a stronger OH absorption. Also, objects without OH detection 
		(shown as black dashed lines) seem to support this idea.
		
		In Figure~\ref{fig:sb}, we present the characteristic velocities measured from the absorption profiles of F00183--7111 
		and 12514+1027 as a function of H-band absolute magnitudes (left panel) and starburst luminosities (right panel). 
		The results are largely consistent with those of V13 on ULIRGs and QSOs: there does not appear to be any strong 
		relation between the molecular outflow velocities and the stellar masses of the hosts or the starburst luminosities (i.e., SFRs). 
		We do note that $v_{50}$ exhibits the smallest dispersion among the velocity measures, and there is a hint of an upward trend 
		between $v_{50}$ and SFR, which the addition of IRAS F00183-7111 appears to dramatically extend. 
		However, it is important to stress that the V13 data do not cover a wide range of SFRs or stellar masses and, unfortunately, 
		our new data do not extend these ranges either. Therefore, one should be cautious when interpreting the lack of 
		obvious correlation between the velocities and SFRs or stellar masses.

		In the left panel of Figure~\ref{fig:agn} we observe that the characteristic velocities of F00183--7111 roughly 
		follow the same correlation with AGN fraction as found among ULIRGs, further suggesting that AGN activity plays a role in driving this outflow. 
		When considering AGN luminosities: in the right panel of Figure~\ref{fig:agn}, the outflow velocities of  F00183--7111
		seem to be consistent within the dispersion with the correlation between outflow velocity and AGN luminosity found by V13, 
		except in the case of $v_{50}$ where F00183--7111 lies beyond the dispersion of the fit.
		
		The fact that OH is observed in emission for IRAS 00397--1312 could indicate it is beginning to shed its dusty cocoon. 
		However, $\alpha_{\rm AGN}$ for the galaxy is only $\sim63$\%, whereas the V13 objects seen with OH purely in emission all had 
		$\alpha_{\rm AGN}\simgt80$\%, which is more consistent with the evolutionary scenario in which the AGN becomes more apparent 
		once the merger remnant has cast off of its natal cocoon and the signatures of AGN feedback are predicted to wane \citep[e.g.][]{N08,H09}. 

                Before jumping to conclusions, one has to remember that scatter in the measured outflow velocities is {\em expected} 
                for several reasons: $(i)$ the measured outflow velocities may be underestimated due to projection effects, which depend on the exact geometry 
                of each outflow; $(ii)$ the kinematics of the outflowing material depend on several largely unconstrained variables (e.g., gas mass fraction and density, 
                column density, dust content of individual cloudlets) associated with the complex multi-phase nature of the entrained ISM and the exact acceleration 
                mechanisms (e.g., radiation pressure, ram pressure, shocks); $(iii)$ we are measuring time-averaged quantities 
                which may not be exactly in phase: the AGN luminosity is derived from mid-infrared measurements on kpc scales while the outflow 
                velocities are from far-infrared measurements on scales of a few hundreds of parsecs (corresponding to a few 10$^6$ yrs for outflow speeds 
                of a few hundred of km s$^{-1}$). Consequently, one cannot draw any firm conclusions based on only two HLIRGs. A larger sample of 
                HLIRGs will be needed for a more meaningful comparison with ULIRGs and QSOs.
		
	\begin{table*}
		\begin{threeparttable}
		\centering
		\caption{Spectral analysis measurements.}
		\begin{tabular}{cccccccc}
			\hline
			\textit{IRAS} Name	&	$v_{50}$(abs)	&	$v_{84}$(abs)	&	$v_{\rm max}$(abs)	&	EW$_{\rm abs}$	&	$v_{50}$(emi)	&	$v_{84}$(emi)	&	EW$_{\rm emi}$\\
			(1)				&	(2)			&	(3)			&	(4)				&	(5)				&	(6)			&	(7)			&	(8)\\
			\hline
			\hline
			F00183--7111	&	$70 \pm 100$	&	$-470 \pm 100$	&	$-1490 \pm 300$	&	$130\pm30$			&	-			&	-			&	-				\\ 
			00397--1312	&	-			&	-			&	-				&	-					&	$220 \pm 60$	&	$590 \pm 200$	&	$-100\pm30$		\\ 
			07380--2342	&	-			&	-			&	-				&	$<250^{\rm a}$ 		&	-			&	-			&	$> -250^{\rm a}$	\\ 
			12514+1027	&	$-50 \pm 60$	&	$-170 \pm 60$	&	$-530 \pm 200$		&	$90\pm10$			&	-			&	-			&	-				\\ 
			14026+4341	&	-			&	-			&	-				&	$< 500^{\rm a}$		&	-			&	-			&	$> -500^{\rm a}$	\\ 
			\hline
		\end{tabular}
		\label{tab:vel}
		\begin{tablenotes}
			\item	\textbf{Notes.} Velocities and EWs are in units of km s$^{-1}$. {\it Column 1:} \textit{IRAS} galaxy names. 
			{\it Column 2:} Median velocity of the absorption profile model fitted. {\it Column 3:} Velocity above which 84\% of the absorption takes place. 
			{\it Column 4:} Maximum extent of the blueshifted wing of the absorption profile. {\it Column 5:} Equivalent width for the absorption components. 
			{\it Column 6:} Median velocity of the emission profile model fitted. {\it Column 7:} Velocity below which 84\% of the emission takes place. 
			{\it Column 8:} Equivalent width for the emission components.
			\item[$^{\rm a}$] The limits appeared in absorption and emission, such that we can only constrain the absolute value of the EW in the spectra with no detection of the OH doublet.
		\end{tablenotes}
		\end{threeparttable}
	\end{table*}

\section{Conclusions}
	\label{sec:conclusions}
	
        We have analysed the \textit{Herschel}/PACS spectrum of five HLIRGs, using the OH119~$\mu$m doublet to search for
        molecular outflows. The OH doublet was detected in three out of five galaxies, from which one was purely in emission and 
        two purely in absorption. Although it is possible that the OH detection in emission may indicate the presence of an outflow, 
        the lack of a clear absorption component in the doublet does not allow us to conclude it with certainty. On the other hand, out of 
        the two systems with OH detections in absorption only IRAS~F00183--7111 seems to have a molecular outflow. The most 
        blueshifted wing of its OH profile reached about 1500~km~s$^{-1}$.
        
        Our analysis supports the lack of correlation between the OH equivalent width and both AGN fraction and luminosity reported by V13. 
        Furthermore, it agrees very well with the correlation between the OH equivalent width and the system obscuration also found by V13. 
        The characteristic velocities of the outflow detected in IRAS~F00183--7111 are in general agreement with previously 
        studied correlations found in studies of local ULIRGs, specifically V13 and \cite{S13}. This fact supports the idea that outflows 
        are not related either to starburst activity or stellar mass. Instead, they support the notion that AGNs are responsible for driving 
        the powerful outflows seen in many ULIRGs/HLIRGs.
        
        A larger sample of HLIRGs is clearly needed to explore these phenomena in more detail. We note that the WISE all-sky catalog is already discovering
        large numbers of new HLIRGs \citep[e.g.][]{E12,T13,To15}. Some fraction of these may lie at suitable redshifts whereby ground-based 
        submillimeter facilities such as ALMA can probe any potential molecular outflows from them. Otherwise, progress may have to wait
        for a still to be defined future far-infrared space observatory.

\section*{Acknowledgments}
	
	We thank the anonymous referee for very useful comments and suggestion to improve this work.
	We acknowledge support from CONICYT-Chile grants, Basal-CATA PFB-06/2007 (DC, FEB, SS), FONDECYT 1141218 (FEB), 
	PCCI 130074 (FEB), ALMA-CONICYT 31100004 (FEB), Gemini-CONICYT 32120003 (FEB), ``EMBIGGEN" Anillo ACT1101 (DC, FEB),
	the Ministry of Economy, Development, Tourism's Millennium Science Initiative through grant IC120009 
	awarded to The Millennium Institute of Astrophysics, MAS (FEB, SS), CONICYT-Chile through PCHA/Doctorado Nacional (2015-21151574) (DC), 
	and by NASA through Herschel contracts 1427277 and 1454738 (SV). Optical observations were obtained with the Magellan/Baade as part 
	of the program CN--2015A--129.

\appendix
\section{Redshift measurements}
\label{ap}
\begin{figure}
		   \centering
		   \includegraphics[width=1.1\linewidth]{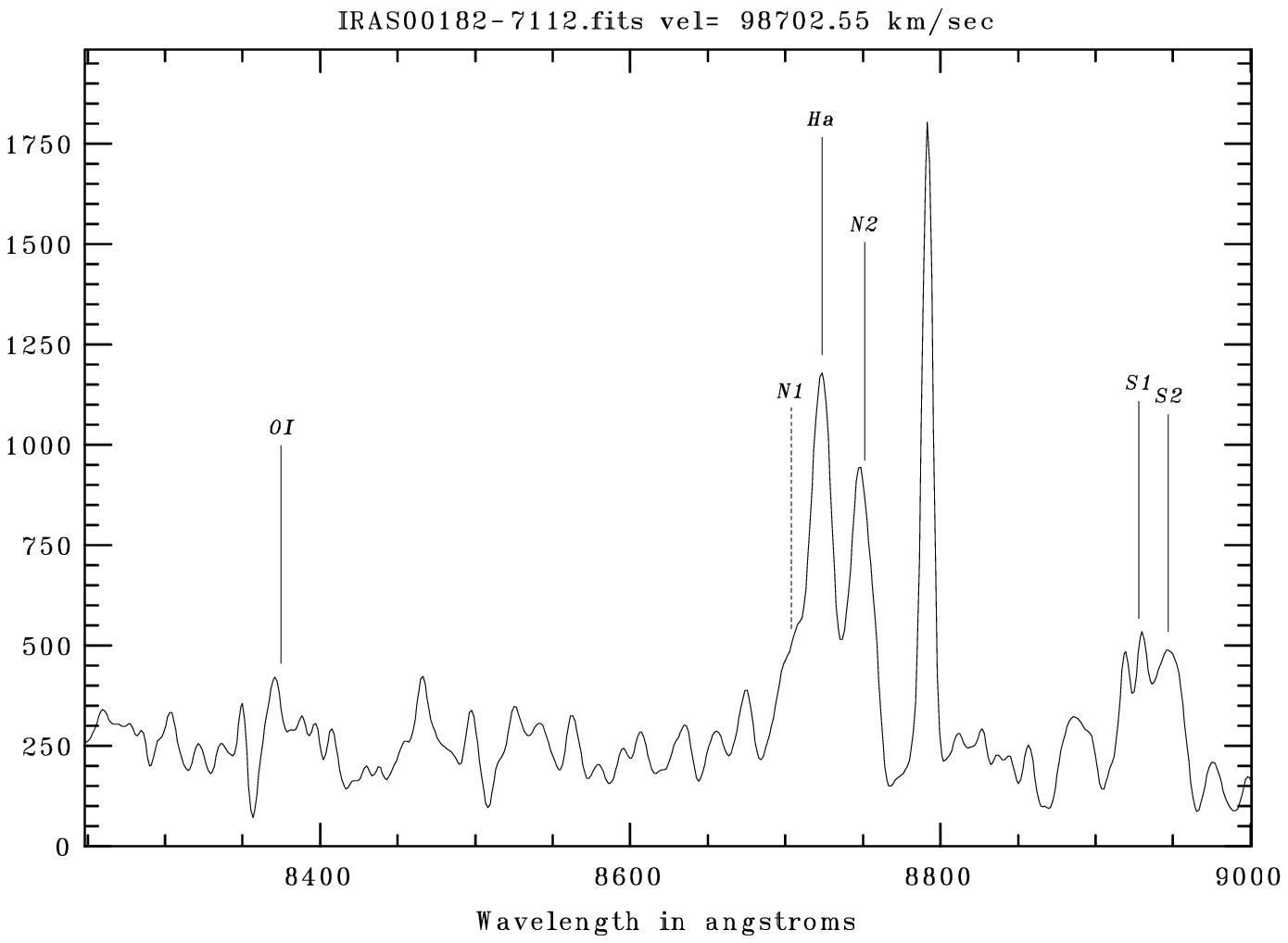} 
		   \caption{Smoothed IRAS F00183-7111 spectrum taken with Magellan/IMACS. The 
		   flux is in instrumental units. The line identification and radial velocity 
		   were performed with the task \textit{emsao} from the \textit{rvsao} package in \textit{IRAF}.}
		   \label{fig:spec1}
	\end{figure}

\begin{figure}
		   \centering
		   \includegraphics[width=1.1\linewidth]{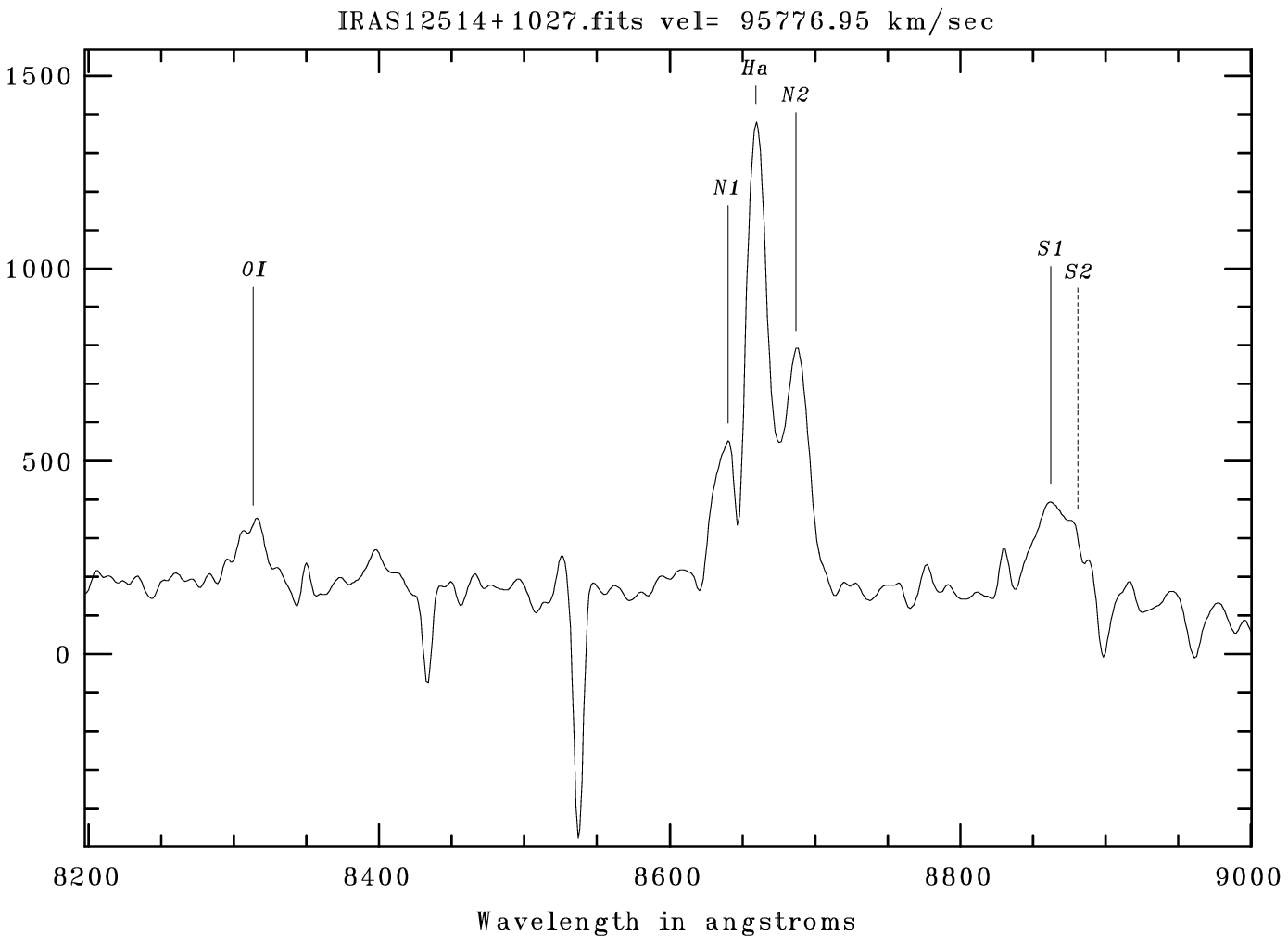} 
		   \caption{Analogous to Figure~\ref{fig:spec1} but for IRAS 12514+1027.}
		   \label{fig:spec2}
	\end{figure}

		Given that systemic velocities are very important in the context of this work, we carried out spectroscopic 
		observations for two galaxies in our sample: IRAS F00183-7111 and IRAS 12514+1027; in order to 
		independently assess the accuracy of their redshifts.
		The spectra were measured using the \textit{Inamori Magellan Areal Camera Spectrograph} (IMACS) on the 
		Magellan I Baade 6.5m telescope at Las Campanas Observatory. We obtained two exposures of 300 seconds 
		for each object. The data reduction was performed using \textit{IRAF} following standard procedures. 
		The spectra presented prominent features in emission that allowed us to estimate its radial velocity easily 
		(see Figure~\ref{fig:spec1} and~\ref{fig:spec2}). To this end, we used the task \textit{emsao} from the package 
		\textit{rvsao} to identify the redshifted emission lines and then to compute a radial velocity value.
		The measured redshifts and their associated errors are presented in Table~\ref{tab:sample}. The errors account 
		for the wavelength calibration solution, the aperture trace and the \textit{emsao} algorithm uncertainties

\bsp

\label{lastpage}

\end{document}